\documentclass[aps, superscriptaddress, showpacs, floatfix, twocolumn]{revtex4-1}
\usepackage{amsmath}
\usepackage{amssymb}
\usepackage{hyperref}
\usepackage{graphicx}
\begin{document}
\title{Entanglement entropy and computational complexity of the Anderson impurity\\ model out of equilibrium II: driven dynamics} 
\author{Zhuoran He}
\affiliation{Department of Physics, Columbia University, New York, New York 10027, USA}
\author{Andrew J. Millis}
\affiliation{Department of Physics, Columbia University, New York, New York 10027, USA}

\begin{abstract}
\vspace*{0\baselineskip}
We study the growth of entanglement entropy and bond dimension with time in density matrix renormalization group simulations of the periodically driven single-impurity Anderson model. The growth of entanglement entropy is found to be related to the ordering of the bath orbitals in the matrix product states of the bath and to the relation of the driving period $T$ to the convergence radius of the Floquet-Magnus expansion. Reordering the bath orbitals in the matrix product state by their Floquet quasi-energy is found to reduce the exponential growth rate of the computation time at intermediate driving periods, suggesting new ways to optimize matrix product state calculations of driven systems. 
\end{abstract}

\maketitle

\section{Introduction}
The control of strongly correlated electron systems via  laser-induced oscillating electric fields is becoming an active area of research \cite{Paoli06,Banerjee18,PhysRevLett.120.123204,doi:10.1080/01411594.2014.971322,1742-6596-500-14-142011}, making it important to study the dynamics of driven strongly correlated models, i.e., evolution of the system with time-dependent model parameters. The Anderson impurity model (AIM) \cite{PhysRev.124.41}, a correlated orbital coupled to a noninteracting bath, is a model of fundamental interest and is important as an auxiliary model in the dynamical mean-field approach \cite{Georges96, RevModPhys.78.865, PhysRevB.88.235106} to correlated electron physics. The development of efficient real-time impurity solvers \cite{PhysRevB.58.5649, PhysRevLett.115.266802, PhysRevB.91.045136, PhysRevB.76.205120} to study the nonequilibrium properties of strongly correlated models such as the Anderson impurity model is a key challenge in this field of research.

We analyze the use of density matrix renormalization group (DMRG) \cite{PhysRevLett.69.2863, RevModPhys.77.259, KLChan2010} as a nonequilibrium impurity solver \cite{PhysRevLett.88.256403, PhysRevLett.93.076401, Wolf14} for the periodically driven AIM, where the system's wave function is represented by a matrix product state (MPS). The key issue in DMRG calculations is the growth of entanglement entropy of the MPS with simulation time. In our previous study \cite{He17} of ``quench" physics (i.e., the evolution following an instantaneous change of interaction and/or hybridization parameters from one set of constant values to another), we found that different arrangements of bath orbitals could dramatically affect the growth of entanglement entropy, and that a particular arrangement (the ``star geometry" \cite{Wolf14}, associated with a proper energy ordering of bath orbitals) led to a very slow (logarithmic) growth of entanglement entropy, enabling simulations of the long-time behavior at a computational cost that grew only polynomially with the simulation time.

In this paper, we study the periodically driven single-impurity Anderson model (SIAM) and find that driven systems are in general more expensive to simulate than quenched systems. There is a critical driving period $T_c$, namely $2\pi$ over the band width of the bath density of states, such that if the driving period $T\!<\!T_c$, the driven system is as easy to simulate as a quenched SIAM, while if $T\!>\!T_c$, the simulation becomes exponentially hard. For driving periods $T\!>\!T_c$, we find that an ordering of the bath orbitals exists, which we call quasi-energy ordering, such that the asymptotic entanglement entropy growth is slow (logarithmic in time, for the noninteracting model, and with a small linear coefficient for the interacting model). However, the initial transient growth of entropy for this bath ordering can be very rapid before the asymptotic limit is reached, which limits the maximum simulation times reachable in practice.

As in our previous work \cite{He17}, we use the 4-MPS scheme developed previously to simulate the real-time dynamics of the driven SIAM. The wave function is represented by a Schmidt decomposition between 4 states of the impurity orbital ($|0\rangle,\left|\uparrow\right>,\left|\downarrow\right>,\left|\uparrow\downarrow\right>$) and 4 bath MPSs correspondingly. The Hamiltonian $H$ of the quenched model is time-independent, while the Hamiltonian $H(t)$ of the driven model oscillates in time periodically.

The rest of the paper is organized as follows. Section \ref{sec:theory} describes the driven SIAM we solve and generalization of our 4-MPS method \cite{He17} to the driven model. In Sec.~\ref{sec:results-nonint}, we present results obtained for the noninteracting SIAM to obtain the asymptotics of entropy growth behaviors in various parameter regimes to scan the complexity diagram. In Sec.~\ref{sec:results-int}, we simulate the interacting SIAM using our method. We will discuss entropy growth and show some physical results of the model. Section \ref{sec:conclusion} is a conclusion and summary.

\section{Theory and method\label{sec:theory} \label{sec:method}}
We consider a single-impurity Anderson model (SIAM) with periodically oscillating model parameters. The most general form of the time-dependent model Hamiltonian $H(t)$ is given by\vspace{-1ex}
\begin{align}
&H(t)=H_d(t)+H_\mathrm{bath}(t)+H_\mathrm{mix}(t),\phantom{\frac{1}{2}}
\label{eq:H-1}\\
&H_d(t)=\sum_\sigma\epsilon_d(t)n_{d\sigma}+U(t)(n_{d\uparrow}\!-\!\textstyle\frac{1}{2})(n_{d\downarrow}\!-\!\frac{1}{2}),
\label{eq:H-2}
\end{align}
\begin{align}
&H_\mathrm{bath}(t)=\sum_{k\sigma}\epsilon_k(t)\,c_{k\sigma}^\dagger c_{k\sigma},
\label{eq:H-3}\\
&H_\mathrm{mix}(t)=\sum_{k\sigma}V_k(t)\,d_\sigma^\dagger c_{k\sigma}+\mathrm{h.c.},
\label{eq:H-4}
\end{align}
where $n_{d\sigma}=d_\sigma^\dagger d_\sigma$ and $\sigma=\,\uparrow,\downarrow$ is the spin label. We go to the interaction picture of $H_0(t)\equiv H_d(t)+H_\mathrm{bath}(t)$. The $H_\mathrm{mix}(t)$ part in the interaction picture becomes
\vspace{-1ex}
\begin{align}
\hat{H}_\mathrm{mix}(t)&=U_0(0,t)\,H_\mathrm{mix}(t)\,U_0(t,0)\phantom{\frac{1}{2}}\nonumber\\
&=\sum_{k\sigma}V_k(t)\hat{d}_\sigma^\dagger(t)\hat{c}_{k\sigma}(t)+\mathrm{h.c.},
\end{align}
where $U_0(t,0)=\mathcal{T}e^{-i\int_0^tH_0(t')dt'}$ is the time-ordered unitary evolution from $0$ to $t$ and $U_0(0,t)=[U_0(t,0)]^\dagger$. Since $H_0(t)$ does not couple the $d$ orbital to the bath, each bath orbital evolves independently in the interaction picture as given by
\begin{subequations}
\begin{align}
\hat{c}_{k\sigma}(t)=c_{k\sigma}\,e^{-i\int_0^t\epsilon_k(t')dt'},
\label{eq:ck-int-pic}
\end{align}
and the $d$ orbital evolves according to
\begin{align}
\hat{d}_\sigma(t)=d_\sigma\,e^{-i\int_0^t[\epsilon_d(t')+U(t')(n_{d\bar{\sigma}}-\frac{1}{2})]dt'},
\label{eq:d-int-pic}
\end{align}
\end{subequations}
with $\bar{\sigma}$ denoting the opposite spin of $\sigma$. Notice that $\hat{n}_{d\bar{\sigma}}(t)=n_{d\bar{\sigma}}$ does not evolve in the interaction picture of $H_0(t)$ and that $n_{d\bar{\sigma}}$ commutes with $d_\sigma$, which together lead to Eq.~\eqref{eq:d-int-pic}. 

As in the 4-MPS scheme developed in our previous work \cite{He17}, the wave function $|\Psi(t)\rangle$ is represented by
\begin{align}
|\Psi(t)\rangle=\sum_i c_i(t)_{\,}|i\rangle_d\otimes|\Psi_i(t)\rangle_\mathrm{bath},
\end{align}
where $i$ sums over the $4$ impurity states $|0\rangle$, $\left|\uparrow\right>$, $\left|\downarrow\right>$, and $\left|\uparrow\downarrow\right>$, and every $|\Psi_i(t)\rangle_\mathrm{bath}$ is a matrix product state. The wave function $|\Psi_i(t)\rangle$ is evolved according to
\begin{align}
|\Psi(t+\Delta t)\rangle&\approx e^{-i\tilde{H}_\mathrm{mix}(t+\frac{\Delta t}{2})\Delta t\!}\,|\Psi(t)\rangle,
\label{eq:evolve}
\end{align}
with the exponential Taylor expanded to 4th order of $\Delta t$ to ensure good unitarity. The time-averaged Hamiltonian $\tilde{H}_\mathrm{mix}(t)$ in a time step $\Delta t$ is now given by
\begin{align}
\tilde{H}_\mathrm{mix}(t)&\equiv\frac{1}{\Delta t}\int_{t-\Delta t/2}^{t+\Delta t/2}\hat{H}_\mathrm{mix}(t')dt'
\nonumber\\
&=\sum_{k\sigma}\tilde{V}_{k\sigma}(t)d_\sigma^\dagger c_{k\sigma}+\mathrm{h.c.},
\label{eq:H-eff}
\end{align}
with effective hopping amplitude given by
\begin{align}
\tilde{V}_{k\sigma}(t)&\approx V_k\,e^{i\int_0^t[\epsilon_d(t')+U(t')(n_{d\bar{\sigma}}-\frac{1}{2})-\epsilon_k(t')]dt'}\nonumber\\
&\quad\times\,\mathrm{sinc}\left(\textstyle\frac{\epsilon_d(t)+U(t)(n_{d\bar{\sigma}}-1/2)-\epsilon_k(t)}{2}\Delta t\right).
\end{align}
The Hamiltonian $\tilde{H}_\mathrm{mix}(t)$ is represented by a matrix product operator (MPO) with bond dimension 2 to act upon the wave function in 4 MPSs \cite{He17}.

\begin{figure}
\includegraphics[width=0.8\columnwidth]{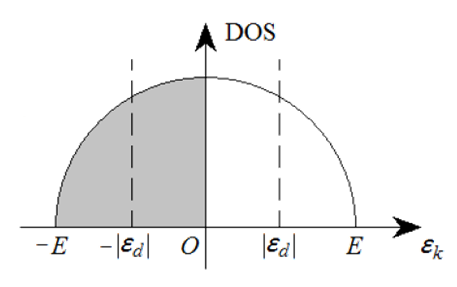}
\caption{The density of states of the bath orbitals $\epsilon_k$. We consider a semicircle DOS with a half band width $E$. The bath is initially half-filled, and the $d$-orbital energy $\epsilon_d=\pm|\epsilon_d|$ oscillates every half driving period $T/2$ across the Fermi level.}
\label{fig:DOS}
\end{figure}

Up to now everything has been general for the single-impurity Anderson model. For concreteness, we consider the evolution starting from a product state
\begin{align}
|\Psi(t=0)\rangle=|\Psi_0\rangle_d\otimes|\mathrm{FS}\rangle_\mathrm{bath},
\label{eq:pure-initial-state}
\end{align}
where $|\mathrm{FS}\rangle_\mathrm{bath}$ is a half-filled Fermi-sea state of the bath with a semicircle density of states (DOS) as shown in Fig.~\ref{fig:DOS}. The $\mathcal{N}\rightarrow\infty$ bath orbitals have fixed energies $\epsilon_k(t)=\epsilon_k$ and fixed equal hopping amplitudes $V_k(t)=V/\sqrt{\mathcal{N}}$ to the impurity $d$ orbital. The Hubbard $U$ on the $d$ orbital is also fixed. The only time-dependent quantity is the $d$-orbital energy
\begin{align}
\epsilon_d(t)=\left\{\begin{array}{ll}
-|\epsilon_d|, & \displaystyle 0<t<\frac{T}{2},\\
+|\epsilon_d|, & \displaystyle \frac{T}{2}<t<T,
\end{array}\right.
\end{align}
which oscillates in a square wave every half driving period $T/2$. A piecewise constant Hamiltonian is numerically easier to handle because Eq.~\eqref{eq:evolve} can be made exact by choosing $\Delta t$ such that $T/2$ is a multiple of $\Delta t$. We will provide physical results which show the local quantities on the $d$ orbital and complexity results which show the growth of entanglement entropy of the bath MPSs at the maximum entropy bond, which is often the one closest to the Fermi level.

\section{Noninteracting Results \label{sec:results-nonint}}
We first do some noninteracting calculations of the driven SIAM using a standard Slater-determinant-based method. In this section, the Hubbard $U=0$ and the initial state is $|0\rangle_d\otimes|\mathrm{FS}\rangle_\mathrm{bath}$, an empty $d$-orbital and a half-filled Fermi sea in Fig.~\ref{fig:DOS}. The impurity-bath coupling $V/E=0.25$. Bath size $N=1000$ orbitals. The time-evolution of a Slater-determinant state by a noninteracting Hamiltonian is numerically cheap and is not limited by the growth of entanglement entropy. The 4-MPS method will be applied in the next section to an interacting SIAM with Hubbard $U>0$.

\subsection{Energy-ordered bath}
\begin{figure}
\includegraphics[width=0.9\columnwidth]{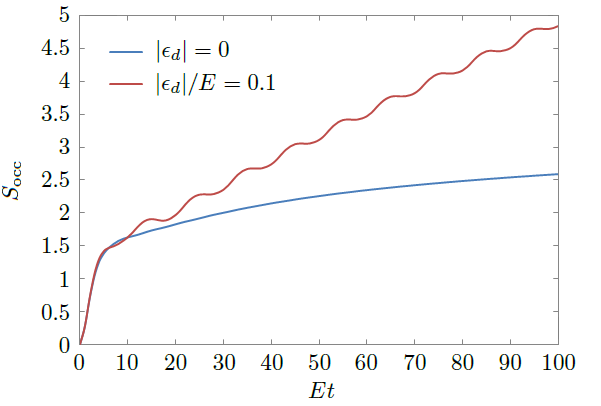}
\caption{(Color online) The entanglement entropy growth of a driven SIAM (red line) against a quench SIAM (blue line). The driving period $ET=10$. Hubbard $U=0$ and impurity-bath coupling $V/E=0.25$. Initially the impurity is empty and the bath is half-filled.
\label{fig:driven-vs-quench}}
\end{figure}

We use the entanglement entropy $S_\mathrm{occ}$ between the $N/2$ bath orbitals below the Fermi level and the rest of the system to estimate the maximum entanglement entropy encountered in an MPS-based simulation when the bath orbitals are energy-ordered. We find that for long driving periods $T>T_c=\pi/E$, since our bath DOS is gapless (see Fig.~\ref{fig:DOS}), an arbitrarily small amplitude $|\epsilon_d|$ changes the logarithmic growth of entropy to linear. This is shown in Fig.~\ref{fig:driven-vs-quench}, where we did a simulation with $N=1000$ bath orbitals and driving period $ET=10$. The entanglement entropy $S_\mathrm{occ}$ between the $500$ bath orbitals below the Fermi level and the rest of the system is plotted in Fig.~\ref{fig:driven-vs-quench} over time. The quenched model exhibits a logarithmic growth of entanglement entropy and the entanglement entropy growth in the periodically driven model is linear in simulation time.

The linear growth of entanglement entropy in a driven SIAM can be intuitively understood in an entropy pumping picture. The up and down motion of the $d$ orbital acts as an elevator that transports some electrons from the occupied bath orbitals to the unoccupied bath orbitals (and holes in the opposite direction). So if the entanglement entropy $S_\mathrm{occ}$ increases by a constant $(\Delta S_\mathrm{occ})_T$ in every period, then the linear growth rate of $S_\mathrm{occ}$ would be $(\Delta S_\mathrm{occ})_T/T$.

\begin{figure}
\includegraphics[width=0.9\columnwidth]{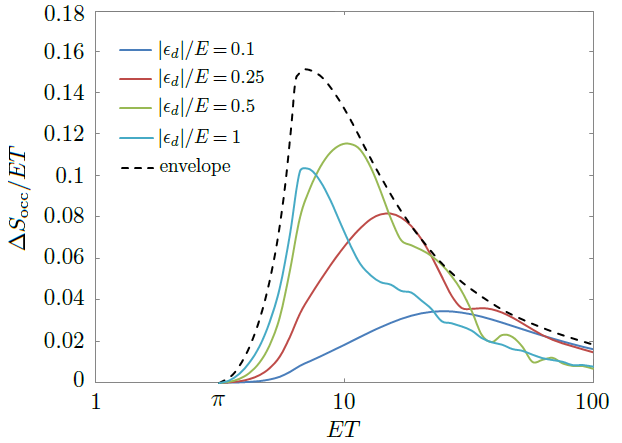}
\caption{(Color online) The steady-state entropy growth rate $(\Delta S_\mathrm{occ})_T/T$ v.s.~the period $T$ at various amplitudes $|\epsilon_d|$. Hubbard $U=0$. Impurity-bath coupling $V/E=0.25$.
\label{fig:critical-period}}
\end{figure}

The critical driving period $T_c=\pi/E$, or $2\pi$ over the band width, separates the logarithmic growth ($T<T_c$) and linear growth ($T>T_c$) of $S_\mathrm{occ}$. Fig.~\ref{fig:critical-period} plots the maximum growth rate of entropy $(\Delta S_\mathrm{occ})_T/ET$ v.s.~the period $T$ as an envelope of the growth rate v.s. $T$ curves at fixed driving amplitudes $|\epsilon_d|$. Each of these curves is tangent to the envelope at some points and they all intersect with zero at the same critical driving period $T_c$. For periods $T<T_c$, the linear growth of entropy cannot be maintained. To understand this critical period, one needs to consider the Floquet Hamiltonian $H_F$ of the driven system defined by
\begin{align}
e^{-iH_FT}\equiv e^{-iH_+T/2}\,e^{-iH_-T/2},
\end{align}
where $H_\pm$ corresponds to $\epsilon_d=\pm|\epsilon_d|$ respectively. The time-independent Floquet Hamiltonian $H_F$ reproduces the unitary evolution of the time-dependent system $H(t)$ over full periods. It turns out $ET=\pi$ is the convergence radius of the Floquet-Magnus expansion of $H_F$ in terms of $H_+$ and $H_-$. Within the convergence radius and for small driving amplitude $|\epsilon_d|$, we have
\begin{align}
H_F=\bar{H}+i|\epsilon_d|\tan\left(\frac{T}{4}\,\mathrm{ad}_{\bar{H}}\right)\!n_d+\mathcal{O}(|\epsilon_d|^2),
\label{eq:H_F-perturb}
\end{align}
where $\bar{H}=(H_++H_-)/2$ is the SIAM Hamiltonian with $\epsilon_d=0$, $\mathrm{ad}_{\bar{H}}=[\bar{H},\cdot\,]$ is the adjoint representation of $\bar{H}$, and $\tan(\cdot)$ is defined via its Taylor expansion. A detailed derivation of Eq.~\eqref{eq:H_F-perturb} is in Appendix \ref{appendix:a}. The convergence radius is $ET=\pi$. We expect this to hold also for the interacting SIAM, because in the thermodynamic limit $N\rightarrow\infty$, the spectral radius $\Vert\bar{H}\Vert$ is mainly determined by the band width of the bath DOS (unless a bound state is formed on the impurity). As a result, $\Vert\bar{H}\Vert\approx E$ is equal to the half band width $E$ of the bath. Since $\tan(\cdot)$ is singular at $\pi/2$, the series expansion of Eq.~\eqref{eq:H_F-perturb} fails to converge if $\Vert\mathrm{ad}_{\bar{H}}\Vert_{\,}T/4=\Vert\bar{H}\Vert_{\,}T/2\approx ET/2>\pi/2$, i.e.,~for long periods $ET>\pi$.

\begin{figure}
\includegraphics[width=0.9\columnwidth]{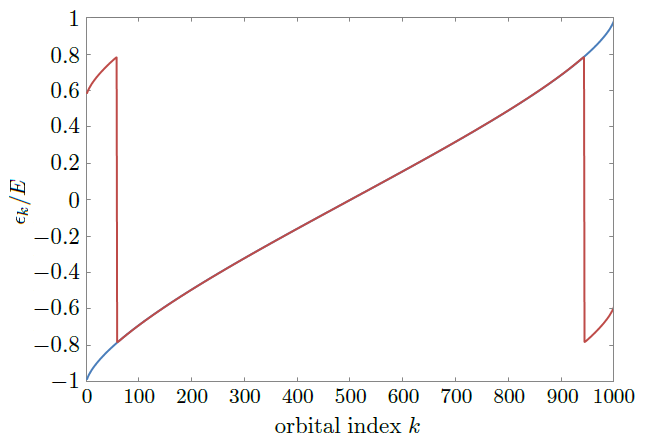}
\caption{(Color online) The bath orbital energies $\epsilon_k$ of the Floquet Hamiltonian $H_F$ for $ET=3$ (blue) and $ET=4$ (red). The orbital energies are unaffected by the periodic driving if $ET<\pi$ but get aliased to quasi-energies within $[-\pi/T,\pi/T]$ modulo $2\pi/T$ if $ET>\pi$.
\label{fig:floquet-energies}}
\end{figure}

Once the perturbative expansion does not converge, surprising new physics emerges. Outside the convergence radius ($T\!>\!T_c$), numerics shows that the bath orbital energies in $H_F$ are aliased to $[-\pi/T,\pi/T]\subset[-E,E]$, breaking the original ordering of the bath orbitals. This situation is shown in Fig.~\ref{fig:floquet-energies}. The inter-bath-orbital hopping amplitudes remain very small. The ascending order of bath orbital energies is violated because of energy aliasing, which gives overlap of energy spectra between the occupied and unoccupied bath orbitals and thus a linear growth of entanglement entropy. 

\subsection{Quasi-energy-ordered bath}

\begin{figure}[b]
\includegraphics[width=0.93\columnwidth]{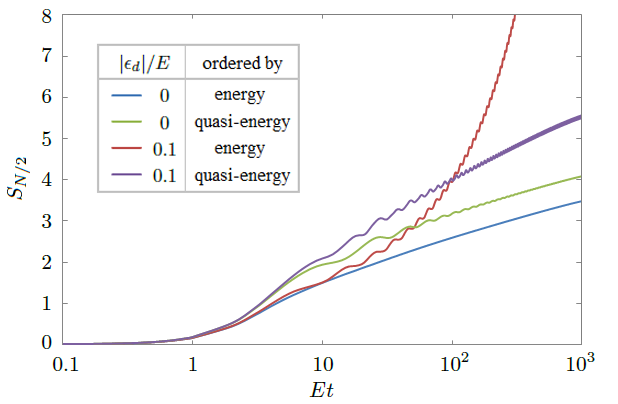}
\caption{(Color online) The growth of entropy $S_{N/2}$ for the driven and quenched models with energy-ordered and quasi-energy-ordered bath orbitals. Hubbard $U=0$ and impurity-bath coupling $V/E=0.25$. Period $ET=10$.
\label{fig:driven-vs-quench2}}
\end{figure}

What happens then if one reorders the bath orbitals in the MPS in ascending order of quasi-energy rather than energy when the driving period $T\!>\!T_c$? We again use the Slater-determinant-based noninteracting simulation to make an estimation. The initial state $|0\rangle_d\otimes|\mathrm{FS}\rangle_\mathrm{bath}$ in the star geometry remains a product state (an MPS with bond dimension $=1$). We use the entanglement entropy $S_{N/2}$ between the $N/2$ bath orbitals with negative quasi-energies (within $[-\pi/T,0)$) and the rest of the system to estimate the maximum entanglement entropy encountered in an MPS-based simulation when the bath orbitals are quasi-energy-ordered. The $S_{N/2}$ defined here becomes equivalent to the entanglement entropy $S_\mathrm{occ}$ used in the previous subsection if $ET<\pi$, when the bath orbitals are energy-ordered.

We redo the same simulation as in Fig.~\ref{fig:driven-vs-quench} using $N=1000$ bath orbitals ordered by their quasi-energies of $ET=10$. The entropies of the energy-ordered simulation in Fig.~\ref{fig:driven-vs-quench} (blue and red lines) are compared with the new results (green and purple lines) in Fig.~\ref{fig:driven-vs-quench2} and the time $t$ is put on log scale. It is found that the growth of $S_{N/2}$ is logarithmic for both the quenched and driven models. This is because the Floquet Hamiltonian $H_F$ is now energy-ordered, as opposed to the aliased situation in Fig.~\ref{fig:floquet-energies}. But the driven model is still harder to simulate than the quenched model, because the slope of the $S_{N/2}$ v.s.~$\ln t$ curve is greater for the driven model.

\begin{figure}
\includegraphics[width=0.9\columnwidth]{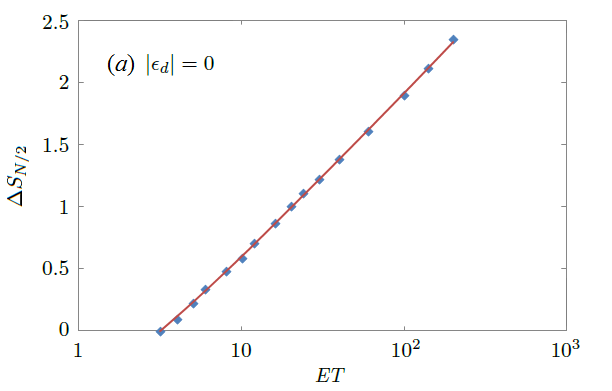}
\includegraphics[width=0.9\columnwidth]{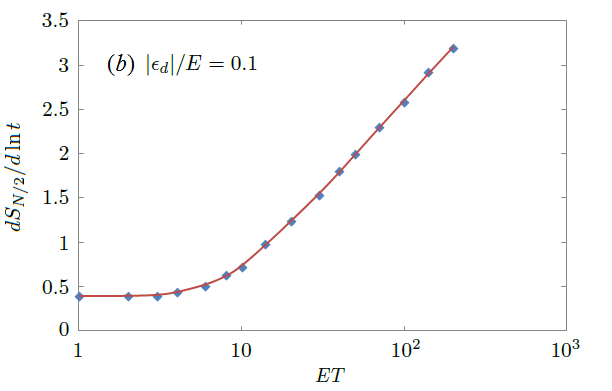}
\caption{(Color online) (a) The upshift $\Delta S_{N/2}$ of entropy in the quenched SIAM at $|\epsilon_d|=0$. (b) The slope of $S_{N/2}$ v.s~$\ln t$ in the driven SIAM at $|\epsilon_d|/E=0.1$. Bath size for long periods need to reach $N=3000$ to obtain accurate data.
\label{fig:nonint-entropy}}
\end{figure}

For the quenched model, the steady-state slope of $S_{N/2}$ v.s.~$\ln t$ is unchanged when the bath orbitals are quasi-energy-ordered. Only the steady-state intercept is shifted up by a constant $\Delta S_{N/2}$, which is found to be approximately proportional to $\ln(T/T_c)$ (see Fig.~\ref{fig:nonint-entropy}a). This is the price to pay for not ordering the quenched bath by energy, which is better than a randomly shuffled bath (see Fig.~7 in \cite{He17}), whose entropy $S_{N/2}$ would grow linearly with time $t$.

\begin{figure}
\includegraphics[width=0.9\columnwidth]{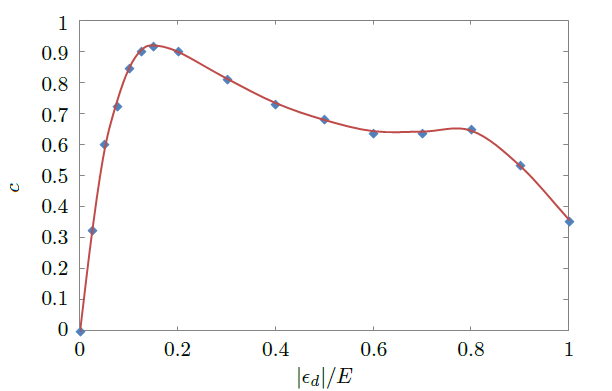}
\caption{(Color online) The dependence of the coefficient $c$ in Eq.~\eqref{eq:entropy-T-t} on the driving amplitude $|\epsilon_d|$. Hubbard $U=0$ and impurity-bath coupling $V/E=0.25$.
\label{fig:c-ed}}
\end{figure}

For the driven model, the driving period $T$ changes the slope of the $S_{N/2}$ v.s.~$\ln t$ curve. Fig.~\ref{fig:nonint-entropy}b shows how the slope increases from that of the quenched model ($T\rightarrow 0$ at fixed $|\epsilon_d|$ is equivalent to quench) to unboundedly large values proportional to $\ln T$. This indicates that the leading-order term in the entropy $S_{N/2}$ is
\begin{align}
S_{N/2}\sim c\ln T\ln t,
\label{eq:entropy-T-t}
\end{align}
where $c$ depends on $|\epsilon_d|$ but is found to be bounded (see Fig.~\ref{fig:c-ed}). At very large $|\epsilon_d|\gtrsim E$, the coefficient $c$ goes down, which is likely to come from the bound state formed on the impurity.

Eq.~\eqref{eq:entropy-T-t} means that the bond dimension in an MPS-based simulation using the quasi-energy-ordered bath arrangement is $D\sim e^{S_{N/2}}\sim t^{\,c\ln T}$. The time complexity of the singular value decomposition (SVD) step is then $\mathcal{O}(D^3)=\mathcal{O}(t^{3c\ln T})$.
Since the power of $t$ for the quasi-energy-ordered method is unbounded for long driving periods $T$, the complexity is still beyond polynomial time. Another drawback of quasi-energy ordering is delocalization of maximum entanglement entropy throughout the MPS, while in energy-ordered MPSs, the maximum entanglement entropy tends to concentrate near the Fermi level. This gives the quasi-energy-ordered method a prefactor of the bath size $N$.

\section{Interacting results \label{sec:results-int}}
In the previous section, we have been estimating what would happen in an MPS-based simulation using a Slater-determinant-based code for the noninteracting SIAM. Now let us do some real MPS-based simulations of the interacting SIAM using the 4-MPS method developed in Sec.~\ref{sec:theory}. We choose a fixed Hubbard $U/E=1$ and the impurity-bath coupling $V/E=0.25$ is the same as in Sec.~\ref{sec:results-nonint}. We use $N=30$ bath orbitals to fit the hybridization function of the continuum bath DOS in Fig.~\ref{fig:DOS} with good accuracy up to $Et\leq 75$ following \cite{He17}. The SVD truncation error tolerance was $10^{-5}$. Noninteracting $d$-occupancies are reproduced with $2\sim 3$ decimal places as a benchmark.

\subsection{Physical results}
\begin{figure}
\includegraphics[width=0.9\columnwidth]{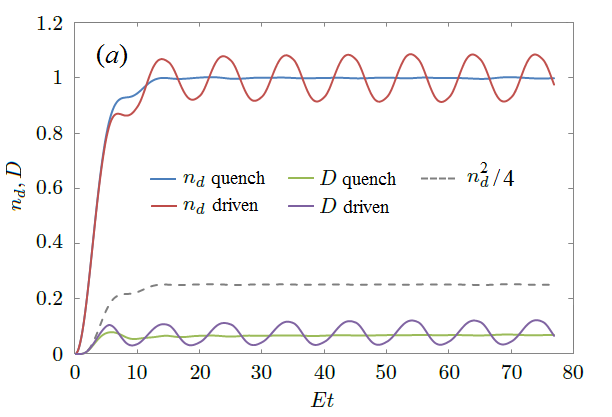}
\includegraphics[width=0.9\columnwidth]{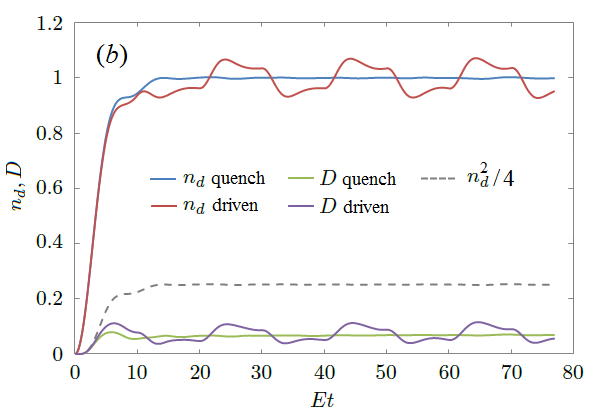}
\caption{(Color online) The $d$-occupancy $n_d=\langle n_{d\uparrow}\rangle+\langle n_{d\downarrow}\rangle$ and double occupancy $D=\langle n_{d\uparrow}n_{d\downarrow}\rangle$ of the quenched and driven SIAMs v.s.~time at Hubbard $U/E=1$, impurity-bath coupling $V/E=0.25$, driving amplitude $|\epsilon_d|/E=0.1$ and period in (a) $ET=10$ and (b) $ET=20$. The dashed grey line is $n_d^2/4$ of the quenched $n_d$.
\label{fig:SIAM-int}}
\end{figure}

\begin{figure}
\includegraphics[width=0.93\columnwidth]{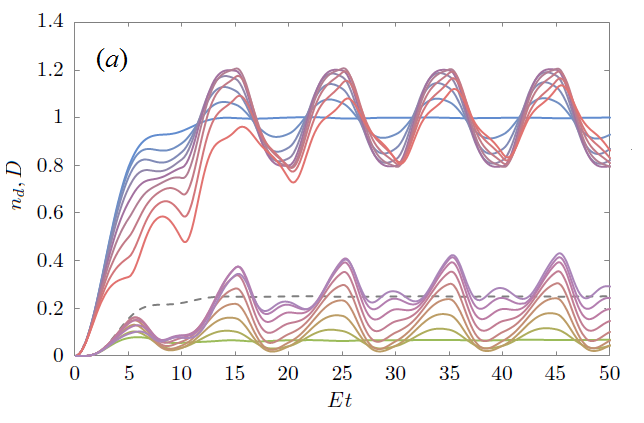}
\includegraphics[width=0.93\columnwidth]{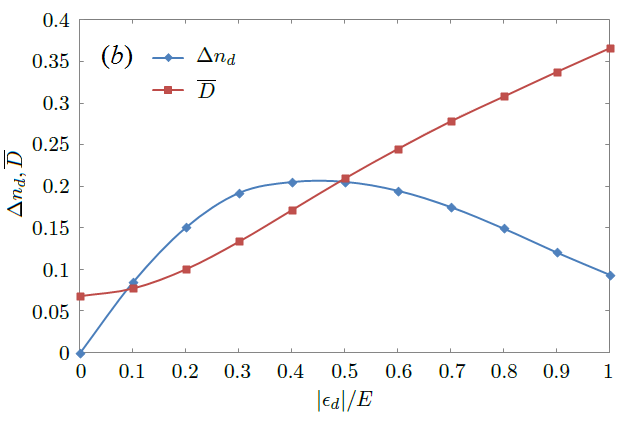}
\caption{(Color online) (a) The $d$-occupancy $n_d$ and double occupancy $D$ of the SIAM at driving amplitudes $|\epsilon_d|/E=0,0.1,\ldots,0.8$ and period $ET=10$. Other parameters are the same as Fig.~\ref{fig:SIAM-int}. The grey dashed line is $n_d^2/4$ of the quenched $n_d$. (b) Amplitude $\Delta n_d$ ($1/2$ of peak-to-peak value) of $n_d$ and the time-averaged double occupancy $\overline{D}$ over a full period.
\label{fig:SIAM-int2}}
\end{figure}

First let us show how the physical results of the interacting SIAM differ from the noninteracting SIAM. The results of short periods $ET<\pi$ are not significantly different from the quenched SIAM with no oscillation of $d$-orbital energy. So we plot both Figs.~\ref{fig:SIAM-int} \& \ref{fig:SIAM-int2} in the long-period regime $ET\!>\!\pi$. Both the energy-ordered and quasi-energy-ordered algorithms as discussed in Sec.~\ref{sec:results-nonint} give the same physical results.

The Hubbard $U$ suppresses the double occupancy of the $d$-orbital for both the quenched and driven SIAMs. In Fig.~\ref{fig:SIAM-int}, the dashed grey lines indicate the level of double occupancy in a noninteracting SIAM (estimated from the $n_d^2/4$ of the quenched $n_d$). The interacting double occupancy is appreciably lower than $n_d^2/4$ when the driving amplitude $|\epsilon_d|/E=0.1$ is small. For period $ET=10$, both $n_d$ (red line in Fig.~\ref{fig:SIAM-int}a) and the double occupancy $D$ (purple line) oscillate in sinusoidal waves, even though the driving signal $\epsilon_d(t)$ is a square wave. When the period increases to $ET=20$, the wave forms approach a relaxed oscillation (Fig.~\ref{fig:SIAM-int}b). The overshoots in every period disappear in a noninteracting simulation ($U=0$, not plotted), which produces simple monotonic decays to the square wave levels.

When the driving amplitude $|\epsilon_d|$ is increased, the wave form of $n_d$ distorts, and the relaxation to steady-state oscillation slows down, as is shown in Fig.~\ref{fig:SIAM-int2}. Also, there is an increase of the average double occupancy $\overline{D}$. At $|\epsilon_d|/E=0.8$, the double occupancy $D$ in its oscillation steady state is above $n_d^2/4$ almost the entire period. A possible explanation might be that the oscillating $d$-orbital energy is like a phonon mode that induces an effective intra-$d$-orbital attraction, which becomes greater than $U$ when the oscillation amplitude $|\epsilon_d|$ is big enough ($|\epsilon_d|/E\gtrsim 0.6$, at which $D\approx 1/4$). Whether this attractive interaction can lead to superconductivity is interesting for further studies.

\subsection{Complexity results}
Obtaining results in Fig.~\ref{fig:SIAM-int2}a at medium to large driving amplitudes was not easy, as they were in the $ET\!>\!\pi$ regime. The linear growth of maximum entanglement entropy makes the maximum bond dimensions in the MPSs increase exponentially with the number of periods simulated. We used some extrapolation techniques to estimate the steady-state quantities in Fig.~\ref{fig:SIAM-int2}b, especially for $|\epsilon_d|/E=0.8$ where the relaxation is slow. Here, in this section we mainly check whether this linear growth of entropy (i.e., exponential difficulty) can be helped by reordering the bath orbitals in the MPSs in\linebreak quasi-energy order.
\begin{figure}
\includegraphics[width=0.95\columnwidth]{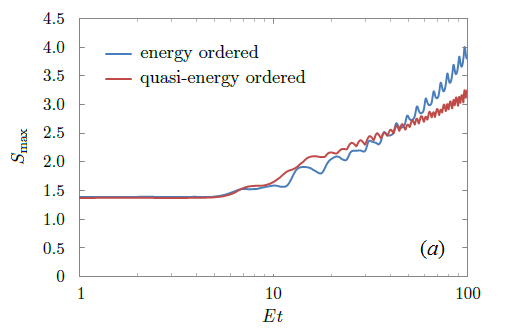}
\includegraphics[width=0.95\columnwidth]{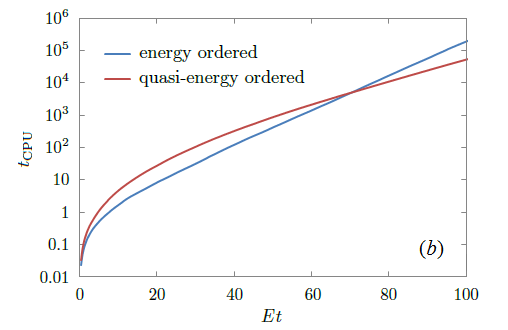}
\caption{The maximum entanglement entropy $S_\mathrm{max}$ reached in (a) and CPU time $t_\mathrm{CPU}$ spent in (b) to run to different simulation times $Et$. Parameter values $U/E=1$, $V/E=0.25$, $|\epsilon_d|/E=0.1$, and $ET=6$. The red curve in (a) is slightly concave upward as $Et$ approaches $100$ when the period-$ET$ oscillations are eliminated by moving average.
\label{fig:E-vs-QE}}
\end{figure}

We find that even though the entropy growth in the noninteracting SIAM changes from linear to logarithmic by quasi-energy ordering the bath orbitals, as is shown in Sec.~\ref{sec:results-nonint}B, the entropy growth for the interacting SIAM is slightly faster than logarithmic (see red curve in Fig.~\ref{fig:E-vs-QE}a). We increase the number of bath orbitals to $N=40$ to reach $Et=100$, and then make a comparison of the energy-ordered and quasi-energy-ordered simulations in Fig.~\ref{fig:E-vs-QE} under $|\epsilon_d|/E=0.1$, $ET=6$. As is shown in Fig.~\ref{fig:E-vs-QE}b, the quasi-energy-ordered 4-MPS simulation is slower than the energy-ordered simulation in the short run. The short-term growth of entropy, e.g.~in the first few periods, is faster if the energies of the bath orbitals are not ordered. In the long run, the quasi-energy ordering is more favorable. The entropy growth only slightly curves up in the $S_\mathrm{max}$ v.s.~$\ln t$ plot. The long-term growth rate of entropy and $\ln t_\mathrm{CPU}$ v.s.~$t$ in Fig.~\ref{fig:E-vs-QE}b are clearly reduced. The hardness in the $ET>\pi$ regime is beyond polynomial time using either method, but is significantly reduced by the quasi-energy ordering method.

\begin{figure}
\includegraphics[width=0.95\columnwidth]{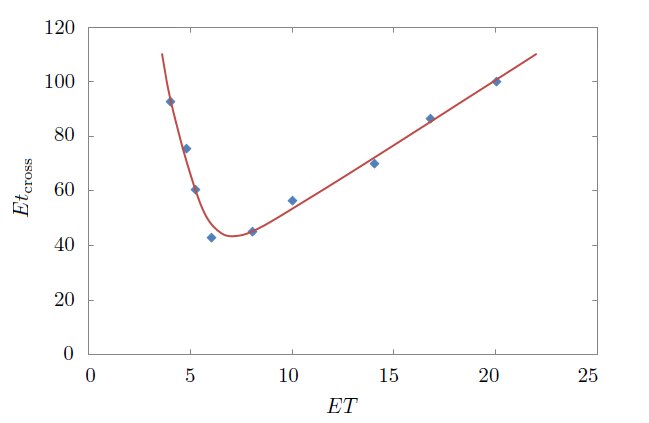}
\caption{Crossing time of maximum entanglement entropies of the energy ordered and quasi-energy ordered simulations at various driving periods $T$. Fixed parameter values $U/E=1$, $V/E=0.25$, $|\epsilon_d|/E=0.1$. The red line is a smooth guideline of the data points in blue dots.
\label{fig:t-cross}}
\end{figure}

Figure \ref{fig:t-cross} shows the crossing time of the maximum entanglement entropies $S_\mathrm{max}$ of the energy ordered and quasi-energy ordered simulations. In a wide range of driving periods the crossing time $t_\mathrm{cross}$ of the entropies in Fig.~\ref{fig:E-vs-QE} exists and is minimum at intermediate driving periods $T$ at which the linear growth rate of entropy $S_\mathrm{max}$ of the energy ordered method is fastest. After the entropies cross, the quasi-energy-ordered method still needs to overcome two more short-term drawbacks: a) its maximum entanglement entropy being more widespread over the MPS bonds than the energy-ordered method, and b) the bigger entropy at short times less than one period, before the actual CPU-times cross.

\section{Conclusion \label{sec:conclusion}}
We have generalized our previously developed 4-MPS method to time-dependent Hamiltonians to study the driven SIAM and have analyzed the computational complexity in the short-period $ET<\pi$ and long-period $ET>\pi$ regimes for both the noninteracting ($U=0$) and interacting ($U>0$) models. The model behavior in the $ET<\pi$ regime is not significantly different from the quenched model. This is the regime in which the Floquet-Magnus expansion converges. Both the interacting and noninteracting models are as easy to simulate as the quenched models (polynomial time). In the $ET>\pi$ regime, the entropy grows linearly in the energy-ordered algorithm, which is therefore exponentially hard to reach long times (many periods). Using quasi-energy ordering reduces the entropy growth of the noninteracting model from linear to logarithmic with a coefficient of the logarithm that grows unboundedly with the driving period $T$ (proportional to $\ln T$). For the interacting model, quasi-energy ordering significantly reduces the linear growth rate of entropy and thus the exponential hardness grows with time more slowly in the long run.

The main challenge of future research is the lack of analytic results for the Floquet Hamiltonian of the interacting model beyond the convergence radius of the Magnus expansion ($ET\!>\!\pi$). Therefore the ``good basis'' of the MPSs has to be guessed. In our work, we made a guess to use quasi-energy ordering based on the noninteracting model. To make better guesses that further reduce entanglement entropy growth, one would need to know the Floquet Hamiltonian of the driven interacting model in greater detail.\\

\noindent
\textbf{Acknowledgments:} This research is supported by the
Department of Energy under grant DE-SC0012375.

\appendix
\section{Floquet Hamiltonian \label{appendix:a}}
The Floquet Hamiltonian $H_F$ of a periodically driven system $H(t)=H_0+\epsilon H_1(t)$ is given by
\begin{align}
e^{-iH_FT}=\mathcal{T}e^{-i\int_0^Tdt\,[H_0+\epsilon H_1(t)]}.
\end{align}
For small amplitudes we have $\epsilon\rightarrow 0$. We can take derivative with respect to $\epsilon$ at $\epsilon=0$ to obtain
\begin{align}
&\mathcal{T}e^{-i\int_0^Tdt\,[H_0+\epsilon H_1(t)]}=e^{-iH_0T}\nonumber\\
&\quad-i\epsilon\int_0^Tdt\,e^{-iH_0(T-t)}H_1(t)e^{-iH_0t}
+\mathcal{O}(\epsilon^2).
\label{eq:a-2}
\end{align}
We define an expansion
\begin{align}
H_F=H_0+\epsilon\,\delta H_F^{(1)}+\mathcal{O}(\epsilon^2).
\end{align}
Then we have
\begin{align}
&e^{-iH_FT}=e^{-i(H_0+\epsilon\,\delta H_F^{(1)})T}+\mathcal{O}(\epsilon^2)
=e^{-iH_0T}\nonumber\\
&\quad-i\epsilon\int_0^Tdt\,e^{-iH_0(T-t)\,}\delta H_F^{(1)}e^{-iH_0t}+\mathcal{O}(\epsilon^2).
\label{eq:a-4}
\end{align}
Comparing Eqs.~\eqref{eq:a-2} and \eqref{eq:a-4}, we have from the first-order terms of $\epsilon$ that
\begin{align}
\int_0^T dt\,e^{iH_0 t}H_1(t)e^{-iH_0t}=\int_0^T dt\,e^{iH_0 t\,}\delta H_F^{(1)}e^{-iH_0t},
\label{eq:a-5}
\end{align}
where we have multiplied on both sides by $e^{iH_0T}$ from the left. Then we use the nested commutator expansion
\begin{align}
e^{iH_0t}H_1(t)e^{-iH_0t}=\sum_{n=0}^\infty\frac{(it)^n}{n!}_{\,}\mathrm{ad}_{H_0}^n[H_1(t)],
\end{align}
where $\mathrm{ad}_{H_0}\equiv[H_0,\cdot]$ is the adjoint representation of $H_0$, and $\mathrm{ad}_{H_0}^n[H_1(t)]=[H_0,\mathrm{ad}_{H_0}^{n-1}[H_1(t)]]$ is the $n$-fold nested commutator of $H_0$ with $H_1(t)$. Using this formula on both sides of Eq.~\eqref{eq:a-5}, and from the square wave model
\begin{align}
H_1(t)=H_1\,\mathrm{sgn}\left(t-\frac{T}{2}\right),\quad 0\leq t<T,
\end{align}
we have
\begin{align}
&\sum_{n=0}^\infty\frac{(iT)^n}{(n+1)!}\left(1-\frac{1}{2^n}\right)\mathrm{ad}_{H_0}^n(H_1)
\nonumber\\
&=\sum_{n=0}^\infty\frac{(iT)^n}{(n+1)!}_{\,}\mathrm{ad}_{H_0}^n(\delta H_F^{(1)}),
\label{eq:a-8}
\end{align}
or in functional form
\begin{align}
\frac{(e^{i\frac{T}{2}\mathrm{ad}_{H_0}}-1)^2}{iT\mathrm{ad}_{H_0}}\,H_1=\frac{e^{iT\mathrm{ad}_{H_0}}-1}{iT\mathrm{ad}_{H_0}}\,\delta H_F^{(1)}.
\end{align}

\pagebreak\noindent
All functions of $\mathrm{ad}_{H_0}$ are defined using their power series in Eq.~\eqref{eq:a-8}. We now apply the multiplicative inverse of the power series of $\mathrm{ad}_{H_0}$ on the right-hand side to both sides and after some algebra obtain
\begin{align}
\delta H_F^{(1)}=i\tan\left(\frac{T}{4}\,\mathrm{ad}_{H_0}\right)\!H_1.
\label{eq:a-10}
\end{align}
The formal solution in Eq.~\eqref{eq:a-10} can be evaluated in the eigenbasis of $H_0$ as
\begin{align}
\langle m|\delta H_F^{(1)}|n\rangle=i\langle m|H_1|n\rangle\tan\left(\frac{E_m-E_n}{4}\,_{\!}T\right).
\label{eq:a-11}
\end{align}
where $|m\rangle$ and $|n\rangle$ are eigenstates of $H_0$ with eigen-energies $E_m$ and $E_n$. In case $H_1=|\epsilon_d|n_d$ with $n_d\equiv\sum_\sigma d_\sigma^\dagger d_\sigma$, Eq.~\eqref{eq:H_F-perturb} in the main text is derived. Since no assumption is made on $H_0$ except it is time independent, Eqs.\eqref{eq:a-10}, \eqref{eq:a-11} \& \eqref{eq:H_F-perturb} hold for both the interacting and the noninteracting SIAMs.

\bibliography{SIAM2_refs}
\end{document}